# Specific Heats $C_p$, $C_v$, $C_{conf}$ and energy landscapes of glassforming liquids.

## C. A. Angell
Dept. of Chemistry and Biochemistry,
Arizona State University,
Tempe AZ 85287-1604
and
## S. Borick
Department of Chemistry
Scottsdale Community College
Scottsdale, AZ 85256-2626

**Abstract**
In pursuit of understanding of the paradoxical success of the Adam-Gibbs equation in both experiment and computer simulation studies, we examine the relation between liquid behavior at constant pressure and constant volume and compare the inherent structures excitation profiles for the two cases. This allows us to extend qualitatively the recent correlation of kinetic and thermodynamic measures of fragility to constant volume systems. The decreased fragility at constant volume is understood in terms of the relation $C_p > C_{c(cp)} > C_{c(cv)} > C_v$. In the process, we find a parallel between the range of volumes, relative to the total excess volume, that are explored in the first few orders of magnitude of relaxation time increase, and the range of amorphous state inherent structure energies, relative to the total range, that are explored in ergodic computer simulations, which also cover only this limited range of relaxation time change. The question of whether or not fragile behavior is determined in the configurational or vibrational manifold of states is left unanswered in this work. However, the approximate proportionality of the configurational and total excess entropies that is needed to interpret the success of the Adam Gibbs equation (which has been questioned by other authors) is confirmed within the needed limits, using data from three different types of investigation: experiments (on Se), simulation (of water in the SPC-E model), and analytical models of both defect crystals and configurationally excited liquids. Some consequences of the abrupt increases in vibrational heat capacity at $T_g$ implied by this proportionality, are discussed.

## 1. Introduction

There has recently been much discussion of the relation between thermodynamic and relaxational properties of glassforming liquids, both from computer simulation studies [1-5] and from laboratory experiments [6-12]. Since the simulation studies have been performed in quite different relaxation time ranges from the experimental studies, and also have been made under different thermodynamic conditions (constant pressure for experiment and constant volume for simulation), the extent to which the findings accord with one another is surprising. In both types of study, the mass transport (diffusion, viscosity) and relaxational properties (from dynamic structure factor, dielectric relaxation time, etc) follow to good approximation the Vogel-Fulcher-Tammann equation

$$\tau = \tau_0 \exp[DT_0/(T-T_0)] \qquad (1)$$

where $\tau_0$, $D$ and $T_0$ are constants. Furthermore, when physical restrictions are placed on the pre-exponent, such that it has a physical value in the vicinity of an inverse attempt frequency of $(10^{13} Hz)^{-1}$ then, in both types of study, the best fit vanishing mobility temperature $T_o$ accords closely with the Kauzmann temperature $T_K$ determined from the vanishing of the excess entropy of the liquid over crystal [3,4,6]. In the case of computer simulation studies, the value of $T_K$ is extracted from purely amorphous state data, by use of a thermodynamic integration from an ideal gas reference state [1-5]. In the case of experiment the $T_K$ estimation relies on an extrapolation referenced to crystal data [6-8]. Even though the crystal reference makes the latter less reliable, for reasons discussed elsewhere [5,10], there has still emerged an equality of the two characteristic temperatures that falls within some ±3% over more than twenty compounds for which suitable data are available [6].

A measure of glassformer fragility that has been recommended by several authors [13, 14] is the ratio $T_o/T_g$. This ratio is inversely related to the parameter $D$ of Eq. (1) which determines the curvature of the Arrhenius plot. Since $T_g$ is an experimentally determined quantity, the equality of $T_o$ and $T_K$ then means that the fragility is the same whether determined from entropy data (i.e. thermodynamics alone) or from relaxation data. Recently [10], a test of this notion of equivalence of thermodynamic and kinetic fragilities was carried out in such a way as to avoid dependence on extrapolated quantities such as $T_o$ and $T_K$. Viscosity data obtained over wide temperature ranges and presented in a $T_g$-scaled Arrhenius plot were compared with the Kauzmann excess entropy [15], also plotted vs $T_g/T$. However the excess entropy was now scaled, not by the excess entropy at the melting point [15], but by the excess entropy at $T_g$. This yields a plot of the same form as the kinetic fragility plot and simplifies the comparison of kinetic and thermodynamic quantities.

By using the $F_{1/2}$ fragility definition [7, 16] and a corresponding quantity in the experimental range for the excess entropy, again a correlation of thermodynamic with kinetic fragility was found [10]. An outstanding exception was the case of $SiO_2$, which is unsettling in view of its status as the archetypal glassformer. However, this problem is a simple consequence of the use of the crystal state as a reference [which is very inappropriate in the case of floppy crystals like cristobalite and trydimite [that have large vibrational entropies at their melting points], and the problem vanishes when the excess entropy is assessed wholly within the amorphous state [5] by the thermodynamic integration method [1-3].

The extent of correlation of thermodynamic and kinetic fragilities found in ref. 10 is very surprising. Although a close correlation between the temperature dependences of free volume, $v_f$, (thermodynamic fragility) and viscosity (kinetic fragility) is predicted by free volume theory [17], the more accepted theory of Adam and Gibbs [18] would lead one to expect major discrepancies. This is because, in the latter theory, the temperature dependence of the relaxation time is expected to be controlled not only by the temperature dependence of the configurational entropy $S_c$, but also by the energy barrier, $\Delta\mu$ per particle, over which the cooperatively rearranging group must pass. $\Delta\mu$ is contained the constant C of the Adam-Gibbs equation,

$$\eta = \eta_0 \exp\{C/TS_c\} \qquad (2)$$

In the free volume model, by contrast, the relation is

$$\eta = \eta_0 \exp(\gamma v^*/v_0) \qquad (3)$$

and, since $v^*$ is expected to be close to the molecular volume [17], the only factor preventing a direct correlation of thermodynamic and kinetic quantities is the free volume overlap factor $\gamma$ [17].

Variations in the $\Delta\mu$ in the AG expression might be expected to be substance-specific, and highly variable [19]. To the extent that the correlation of ref. 10 is real, the finding implies a great simplification in the topography of the energy landscape. Interbasin barriers (saddle points) must scale in some simple manner with the basin energies and populations. Some reasons why this might be the case have been provided recently by Wales [20].

By determining the probability distribution of energy minima (basins of attraction [21]) on the configuration space energy landscape, Sciortino and co-workers [2,4,5] and Sastry [3] have demonstrated that the form of Eq. 1 is valid for a number of systems, insofar as properties like the diffusivity and relaxation time for density fluctuations [4] are linear functions of $\exp(TS_c)^{-1}$. Separately, experimentalists [7,22-24] have tested Eq. (2) successfully for temperatures not too far from the $T_g$, using the excess entropy of Kauzmann's plot (i.e. of the ref. [10] correlation) in place of $S_c$, for lack of alternatives. Again, linear log D vs $1/TS_c$ plots have resulted. Since it is known from the analyses of Goldstein [25], and Goldstein and Gjurati [26] (recently extended by Johari [27]), that the two quantities are not the same, this result has also been surprising. However, Ref. 10 pointed out that the Eq. (2) would be valid for both $S_c$ and $S_{ex}$ if the difference between the two were to originate, in mutual consequence of the elementary configurational excitations, at the Kauzmann temperature. Above $T_K$, due to the extra vibrational entropy generated in the excitations, the two quantities increase at steady but different rates, as depicted in Fig. 1. Then, to first approximation,

$$S_{ex} = a_1(T-T_K), \quad S_c = a_2(T-T_K), \quad S_{ex}/S_c = a_1/a_2 = A, \qquad (4)$$

and Eq. (2) would be valid in the form,

$$\eta = \eta_0 \exp\{C'/TS_{ex}\} \qquad (5)$$

with $C' = C/A$.

Fortunately, there are data available from each of (i) experiments, (ii) simulations, and (iii) simple analytical models which make it possible to verify, within the limits needed to understand the experimental findings, that the approximate proportionality of $S_{ex}$ and $S_c$ is real. This important matter and its consequences will be dealt with in detail in Sections 5 and 6 of this paper. The experimental equivalent of Fig. 1, may be seen reproduced from the work of Phillips et al [28] in this volume, in the paper by Johari [29].

Anticipating the results of section 5 we note here that, at constant pressure, the proportionality constant between $S_{ex}$ and $S_c$ ranges from about 2 for fragile liquids to 1 for strong liquids. Furthermore, according to published data [3], the proportionality constant must continue to fractional values for certain fragile liquids at constant volume.

The implication is that, in general, the relation $C_p > C_{c(p)} > C_{c(v)} > C_v$
(where $C_c$ is a configurational component of the total heat capacity) should hold, as has already been pointed out [10].

To see how this must be, we turn to the consideration of the thermodynamic fragility as it is demonstrated by the landscape "excitation profiles" currently being determined from computer simulations and experiment.

## 2. Energy landscape "excitation profiles" for constant pressure vs constant volume systems, and the thermodynamic fragility of liquids.

According to Goldstein [31] and Stillinger [21,31, 32], the energy landscape for a system of interacting particles at constant volume is uniquely determined by the intermolecular potential function. The details of this multidimensional landscape will depend on the volume held constant. It is evident from the existence in some systems of polyamorphic forms, with very different physical properties, that the landscapes at different volumes can be very different indeed [33].

Most experiments are conducted at constant pressure, and most representations of system behavior, both tabular and graphical, are made for a constant pressure of one atmosphere. Depending on the system under consideration (whether it has a large expansion coefficient or not) the recorded behavior may involve a considerable range of volumes. To utilize the energy landscape concept to interpret the behavior of such systems, it is necessary to consider the case in which an additional dimension is added to the usual constant volume landscape of 3N+1 dimensions, in order to accommodate the volume variable [32].

While the single extra dimension may seem trivial against the total of 3N+2, it makes an important difference to how we should represent the landscape topology in the simple two dimensional multi-minima diagrams often used in discussion of the subject of landscapes. For instance, Sastry's study of the chemically ordered binary Lennard-Jones system at constant volume [3b] has made clear that, when occupying the basins at high energy on the landscape, systems possess higher average vibrational frequencies, hence should be depicted as narrower (sharper) than the lower energy basins. In constant pressure systems, on the other hand this will not be the case, and the basins at high energy (more correctly, high enthalpy though this distinction is not important near zero pressure) should be depicted as shallower [10]. (This may be true even at constant volume when the low temperature structure is one of higher volume, as shown by Starr et al [34]). For the BLJ system, the difference is not large, as can be seen from data on densities of states for different volumes in the study of the constant pressure BLJ system by Vollmayr et al [35].

An important thermodynamic consequence of this difference is that the system at constant pressure will ascend its energy landscape more rapidly at constant pressure than it will at constant volume. This is because the extra vibrational entropy available at higher energies provides as extra entropic driving force to occupy the higher energy levels when temperature is increased. This is an important component of the higher heat capacity of constant pressure liquid systems over constant volume liquid systems. The thermodynamic relation is

$$C_p = C_v + VT\, \alpha/\kappa_T \qquad (6)$$

where $\alpha$ is the isobaric expansivity and $\kappa_T$ is the isothermal compressibility. Although Eq. (6) is written for the total heat capacity, the relation for the configurational component of the heat capacity will have a similar form.

**Excitation profiles**

From the above considerations, the "excitation profile" [36] at constant pressure must be steeper than at constant volume. The excitation profile is the plot, vs T, of the inherent structure energies sampled most frequently by the fluctuating (equilibrated) system. It represents the rate (per K) at which the system climbs its energy landscape as temperature is increased.

This profile was first displayed in detail by Sastry et al [37] who described the fixed volume behavior of the chemically ordered binary mixture of Lennard-Jones particles that is now the model system for studies of this type. This mixture, in which the mutual attractions have been chosen so as to maximally stabilize the liquid without promoting stability of a new crystalline structure, has resisted crystallization during all studies conducted so far. The ergodic part of the excitation profile, obtained from cooling at different rates, and sampling periodically to determine, by conjugate gradient quenching, the inherent structure at that temperature, is reproduced in Fig. 2. The non-ergodic part, where the system fails to equilibrate in the time allowed by the cooling, is included in Fig. 2 only for the case of the most slowly cooled sample. This case is represented by open triangles.

For infinitely slow cooling the behavior is not known in detail though the limiting energy itself can be assumed to be that reached at the Kauzmann temperature, 0.20 on the reduced energy scale of Fig. 2 [37, 1,4]. The curvilinear approach to the ground state shown in Fig. 2, dotted line, is that suggested by simple "excitations" or "defect" models of the thermodynamics of glassformer systems [36,38] which have a binomial density of configurational states. The abrupt, almost singular, approach to the ground state, shown as a dashed line, is that anticipated by models that assume a Gaussian density of configurational states. The latter is the assumption of the random energy model of Derrida [39], and also of the more detailed Random First Order Transition model of Xia and Wolynes [39]. The latter models, which both imply that the ideal glass transition is a singularity, or nearly so, are currently preferred because the rounded maxima in heat capacity, at temperatures above $T_g$, predicted by the defect models are not found [36, 38] except in strong liquids near their strong-to-fragile transition temperatures [5, 41].

Included in Fig. 2 are the changes that must be anticipated for the same system studied under different fixed volumes. The additional profiles shown in Fig. 2 are for volumes that are successively larger than that studied in ref. [37], and which therefore allow the system to reach energies successively closer to the reference value of the dilute gas, for which E = 0.

The isochoric plots at higher volumes have the peculiar feature that they terminate suddenly without entering the glassy state. This is because the higher volume isochores enter the negative pressure domain and intersect the gas liquid spinodal before they become structurally arrested in the glassy state [3]. The closest laboratory parallel of this behavior is the experimentally documented behavior of water (in inclusions in quartz crystals) along isochores extending into the negative pressure regime [42]. We are not concerned here with this aspect of the system's isochoric behavior, though it is good to be aware of its existence. While our adaption of this excitation profile to different volume is qualitative, some of the data needed for a quantitative version are available [3].

The thick line with the steeper slope in Fig. 2 is the locus of points along which the pressure remains constant. It has, of course, a steeper slope than the constant volume excitation profile. It is clear, also, that the Kauzmann temperature must be higher for the depicted constant pressure system.

**Thermodynamic fragility at constant volume vs constant pressure**

It is obvious in Fig. 2 that, according to the thermodynamic criterion, the behavior of a system at constant pressure is much more fragile in character than that of the same system at constant volume. The system climbs to high inherent structure energies much more rapidly at constant pressure than it does at constant volume. The slope difference depicted in Fig. 2 is a factor of 2, which is not unrealistic for a fragile liquid. At least a part of the reason for this has already been given in the inserts to Fig. 2 based on Sastry's findings [3b]. Sastry showed that the shapes of the basins in which the system was trapped during conjugate gradient quenching from high temperatures is different and sharper than for basins explored at lower temperature. This means that the vibrational entropy of the system at high temperatures is not as high as would be expected for the same system confined to any single one of its low temperature basins. Another way of saying this is that for isochoric systems,

$$\Delta C_{v\,(ex,\,vib)} < 0 \qquad\qquad\qquad (7a)$$

According to data in ref. 3b this effect is more important the larger the volume. At constant pressure however, the situation is reversed. There the shallower basins encountered at large volumes [10,35] make

$$\Delta C_{v\,(ex,\,vib)} > 0 \qquad (7b)$$

As we have described elsewhere [10], there is an implication here that an important source of fragility differences between constant pressure liquids lies in differences in the excess vibrational heat capacity. However, this is subject to experimental confirmation. It may be that the extra vibrational heat capacity driving the system to the top of the landscape is a small effect relative to that coming from density of configurational states differences [3b].

The difference between $C_{p,ex}$ and $C_c$ could also arise from anharmonic effects. Indeed the latter was the preferred conclusion of Goldstein, based on his analyses of this matter for constant pressure systems in the '80's [25]. Anharmonicity was also the basis of early attempts by one of us to interpret the origin of the glass transition itself, and link it to the Debye temperature for the glass [43], a relation that should be more accurate for the more fragile, less harmonic, liquids. It will take more analyses of the type carried out by Phillips et al [28], and Suck [44] on systems of widely varying fragility, before this interesting question of the primary source of fragility in liquids at constant pressure (vibrational frequencies vs vibrational anharmonicity [36,10] vs configurational density of states [3b]), can be resolved. Regardless of the molecular level origin of fragility, though, the fragility of a system at constant volume must be less than that of the same system at constant pressure.

**3. Kinetic Fragility at constant volume vs constant pressure.**

In view of the reduced heat capacity at constant volume, the Adam-Gibbs equation, Eq. (2), predicts that the kinetic fragility should be smaller at constant volume, provided that the energy barrier term does not, for some reason, begin to dominate. For experimental systems there is very little information available on which to base a discussion of this question. While a number of high pressure studies have been performed [45,46] there has until very recently [47] been little done on the isochoric temperature dependence of viscosity since the classic but limited study of Jobling and Lawrence [48]. The results of Jobling and Lawrence on n-hexane are reproduced in Fig. 4 by way of counterbalance to the results on triphenyl phosfite and glycerol in ref. 47 because they show behavior that should be observable in simulated systems.

While the data of Fig. 4 are insufficient to show that viscosity still departs from Arrhenius behavior at constant volume [47] they are valuable for showing how the activation energy for viscosity can becomes dramatically smaller at constant volume than it is at constant pressure, and may even vanish for large values of the volume. This dominance of volume effects over temperature effects seen in large-volume samples would seem, at first sight, to run counter to the dominance of temperature asserted in ref 47 as a variable for glassforming systems. However the large-volume systems are not glassforming even if n-hexane is replaced by the strongly glassforming isomer 3-methyl pentane. This is because the large volume samples, like those seen in Fig. 2 of this paper, and in ref. 3a, will cavitate before they vitrify. For the lower-volume glassforming systems it is indeed temperature, not volume, that is the important variable.

From the locus of constant pressure points in Fig. 3, it can be seen that the viscosity behavior must be much more fragile at constant pressure than at constant volume, at all but the smallest volumes. Behavior comparable to that seen in Fig. 3, for viscosity, and also for dielectric relaxation in ref. 48, has been seen for the electrical conductivity in the fragile ionic model system CKN ($4Ca(NO_3)_2 \cdot 6KNO_3$).[49, 50]. These data, at least confirm the simulation finding [3,35] that the transport behavior remains in systematic violation of the Arrhenius law as the temperature is changed.

The data on CKN, are quite old [49], but have not previously been given the attention they deserve. They are therefore reproduced in Fig. 4 (a) and analyzed further. The data for the lower of the two curves (marked 58.8 cc/mole) are obtained by combining the data on conductivity vs pressure in the range 1-1500 atm pressure reported in ref. 49(b) with the data on compressibilities given in ref. 49(c). It is clear that the *average* activation energy energy is lower at constant volume than at constant pressure, as seen in Fig. 3. I

In Fig. 4, however the data are precise enough to ananlyze using Eq. (1), and the parameters obtained allow us to compare the fragilities under the two conditions. The best fit pre-exponential constant, interestingly enough, is the same in each case. The parameters D and $T_o$ found are 1.77 and 320.4K, and 2.50 and 302K, for constant pressure and constant volume, respectively. While $T_g$ for this volume was not directly measured, meaning $T_g/T_o$ ratios cannot be cited, they are proportional to the parameter D (the "strength" parameter), so it is clear that the liquid "strength" is approximately 40% higher at constant volume than at constant pressure for this case.

In Fig. 4(b) the reciprocal of the constant volume activation energy, evaluated using the relation

$$E_v = E_p - \pi \Delta V = E_p - T \Delta V \alpha / \kappa_T \qquad (8)$$

(where $\pi$ is the internal pressure $(\partial P/\partial T)_v$, and $\Delta V$ is the activation volume defined by $\Delta V = RT \partial (\log[\sigma V_{MJ}]/\partial P)$ and $\alpha$ and $\kappa_T$ were defined for Eq. (6)) is shown as a function of the volume held constant. The linear relation implies a value of the volume at which the activation energy diverges and the ideal glass state is encountered, as at $T_K$ of Fig. 2. What is striking about Fig 4(b) is that, in the range in which the conductivity changes by some four orders of magnitude, insufficient to reach the crossover temperature [7,51], the volume has changed over 2/3 of the total change needed to reach the limiting value. Referring to Fig. 2, one notes the parallel between the latter behavior and the range of inherent structure energies explored by the constant volume BLJ system, as its relaxation time increases over approximately the same range.

## 4. Experimental quantification of the excitation profile.

The experimental counterpart of the landscape quantification seen in Fig. 2 is only just being undertaken [51, 52]. The approach involves the use of ultrafast quenching techniques. The potential energy of a glassformer at different temperatures above $T_g$, can be determined if the liquid can be quenched at different rates and the excess energy of the quenched glass, over that of a standard glass, then determined in a "recovery" scan. The "standard glass" is one obtained by some standard cooling process, a convenient choice being 10K/min.

During cooling at a rate Q K/min, a glass is always trapped in a state in which its relaxation time at the point of trapping is

$$\log (t/s) = \log (Q/Ks^{-1}) + 1.5 \qquad (9)$$

This relation is consistent with the observation that the standard glass transition occurs at a temperature where the relaxation time is ~100s [53-55]. Here the standard glass transition temperature is that defined by any of (a) the midpoint cooling glass transition for 10K/min cooling [56] (b) the onset glass transition temperature for a glass scanned at 10K/min after cooling at 10K/min, and (c) the fictive temperature deduced by the Moynihan construction for a glass cooled and heated at 10K/min. These temperatures are found to all be the same within 0.5K for systems of widely varying fragility [51,57].

It appears that, with the fastest possible laboratory quenches, trapped-in states approaching those of the crossover temperature, where the α–β bifurcation occurs, might be possible. Thus the upper limits of laboratory exploration of the landscape may overlap the lower limits of computer simulation. To date the findings are consistent with the existence of shallower traps at higher energies [51]. Details can be obtained by a combination of aging and recovery scan experiments [52]. There are suggestions of interesting structures in the density of configurational states of strong glassformers waiting to be evaluated.

## 5. Concerning the proportionality of $S_{ex}$ and $S_c$

With the above background on energy landscapes in place, we now return to the relation between the total excess entropy of the liquid and the component $S_c$ which is configurational in nature, in the search for a resolution of the Adam-Gibbs equation paradox, i.e. that the equation describes the behavior equally well when total excess entropy is used, in place of configurational entropy alone, in the data fitting. The resolution of the paradox will lie in the demonstration that, in the observable range (i.e. above $T_g$ for experimental systems, or above $T_c$ for simulated systems), the total excess entropy is proportional to the configurational entropy as discussed in the introduction. Since two different authors [29,58] have stated or implied that such proportionality does not exist, it is important that we examine this matter carefully. Thus we take three different approaches to the problem, addressing relevant data from (a) laboratory experiment, (b) computer simulation and (c) analytical modelling.

We show that in each case the proportionality does exist in the relevant range of data, to well within the precision needed to resolve the paradox. On the other hand, it needs to be noted that the tests of the AG equation using alternative entropy quantities have not so far been conducted on the same substance in the same relaxation time range. In the case of the experiments on fragile liquids the Adam-Gibbs equation using $S_{ex}$ only linearizes the data up to the crossover temperature, $T_x$ [19,59] (where $T_x$ is usually found to be the same as $T_c$ of the MCT and $T_b$ of the Stickel analysis [7]). On the other hand, in simulation the AG equation can only be tested for data obtained above $T_x$ for reasons of computational limitations. In the simulation case, both entropy quantities are available, and it will be clear from the case described below, that either quantity would linearize the data. The values of the constants C obtained from the linear plots have not yet been compared with theoretical expectations in either case.

**(a)** *Experiment*:

The only system for which there exists data suitable for direct assessment of the separate vibrational and configurational components of the excess entropy of liquid over crystal, is selenium. Earlier we referred to the data of Phillips et al [28] as the experimental counterpart of Fig. 1. Here we use the data to make a quantitative check of the proportionality of $S_{ex}$ and $S_c$ for this system. For this purpose we use the upper section of Fig. 3 of ref. 28 which is reproduced in this volume in the article by Johari [29]. Johari points out that the entropy of fusion of Se is incorrectly represented in this figure and that an additive correction of 0.43 units of R is required. After adding 0.43 units of R to the total liquid entropy curve above $T_g$ we subtract the crystal entropy to obtain $S_{ex}$. We run a smooth curve through the experimental points for the vibrational heat capacity of the liquid, and measure the difference between this line and the total entropy in order to obtain the configurational component of the excess entropy, $S_c$. These quantities assessed at 50K intervals are plotted in Fig. 5. The ratio $S_c/S_{ex}$ is also plotted using large solid circle symbols. As these assessments were made by hand, using a rule on an enlargement of the figure, they are subject to reading error, so a separate more error-prone assessment of $S_c$ was made by measuring the vibrational excess entropy over crystal first and then subtracting that quantity from $S_{ex}$. This second assessment of $S_c$ is also plotted and the ratio to $S_{ex}$ is represented at the same five temperatures by large open squares**.** The solid lines through the points are best fit 3$^{rd}$ order polynomials. It is noted that, except for $S_{ex,vib}$, they yield the same Kauzmann temperature that was assessed for Se in ref. 60, viz., 230 ± 10K.

The dashed horizontal line is for $S_c/S_{ex} = 0.68$ exactly. Within the measuring "noise", the proportionality of $S_{ex}$ and $S_c$ is confirmed by the two sets of $S_c/S_{ex}$. It is clear that the linearity of Eq. (2) fits of data for selenium would not depend on the choice of $S_{ex}$ or $S_c$ for the fitting. The two open triangles are numerical values quoted in ref. 29, and they provide the basis for the conclusion of that paper that $S_{ex}$ and $S_c$ are *not* proportional.

While discussing the Kauzmann temperature for Se, it is worth noting that the $S_{ex}$ data of Fig. 5 can be equally well fitted using a two parameter equation from the elementary excitations model for glassformer thermodynamics [38(d), 61] without requiring a Kauzmann singularity. It is only in the derivative plots, $C_p$ vs T for selenium and other liquids [38(d), 61] supported by theoretical arguments for a Gaussian density of configurational states in glassformers [1-5,39,40], that evidence favoring the latter can be found.

Speedy [58] has devised a method of assessing the separate contributions of $S_{ex, vib}$ and $S_c$ to the total $S_{ex}$ of ethyl benzene. He assumes that $S_c$ must go to zero before the Kauzmann temperature, so obviously $S_c$ and $S_{ex}$ cannot be proportional near $T_K$. However, above $1.15T_K$ (hence above $T_g$), the ratio $S_c/S_{ex}$ has recovered from the effect of that assumption and at higher temperatures remains constant, within 5%, at the value 0.41, cf. 0.68 for selenium. Ethyl benzene is a very fragile liquid whereas selenium is intermediate (except very close to $T_g$ [**36**]), so the difference is not unexpected. Some consequences of the implication that in fragile liquids the excess entropy of the liquid may be more than half vibrational in character, will be examined briefly in a later section.

**(b)** *Simulation:*

In the study of water in the SPC-E potential [2,62] it was reported that the shapes of the configuration space basins sampled by the system as temperature rises above the crossover temperature, change rapidly in the manner depicted in Fig.1 even though the system is being studied at a fixed volume. It can therefore be used as a bridge between constant volume and constant pressure cases.

This system has been evaluated carefully to distinguish the true minima from the shoulder states, so the configurational entropy has been well quantified. Also the total entropy was evaluated in the course of determining the Kauzmann temperatures for this system [2], so both the quantities needed to test the proposed proportionality are available. While their relation was not investigated in the initial publications [2, 61] they have since been presented both as independent quantities varying with temperature, and in relation to one another [12]. The findings are reproduced here, in Fig. 6, and it is seen that the $S_c$ and $S_{ex}$ are indeed proportional, to very good approximation, despite their individual complexity. The proportionality constant is 0.77 (somewhat above the value for Se), so A of Eq. (4) is 1.3. This is essentially an "experimental" result on a model system studied at constant volume.

**(c)** *analytical models:*
*(i) crystals*.

In this section we will first show that the standard model for thermodynamics of defect crystals also predicts $S_c/S_{ex}$ to be constant over a wide range of defect concentrations. This is because the theory contains an expression for the excess entropy which is different from the configurational entropy for any case in which the creation of a defect is accompanied by a change in lattice frequencies in the vicinity of a defect. The configurational entropy in these models is fixed by the assumption that defects are equally likely to locate on any of the lattice sites, meaning the density of configurational states is binomial, not Gaussian as suggested for liquids.

For a crystal system in which defects can form at the cost of enthalpy ΔH per mole, the number of intrinsic defects, n, present at temperature T in a lattice of N sites, is known [63] to be

$$n = (N-n) \exp(-\Delta H/RT). \qquad (10)$$

Since, in crystals, defects only occur in numbers that are small compared to the number of lattice sites, this expression is usually subjected to the approximation n/(N-n) ≈ n/N. and the standard exponential relation between defect fraction and temperature results, viz.,

$$n/N = \exp(-\Delta H/RT). \qquad (11)$$

However, if many defects can form, then this approximation cannot be tolerated, and instead the defect population n/N must be written, by rearrangement of Eq (10), as

$$n/N = [1 + \exp(\Delta H/RT)]^{-1} \qquad (12)$$

This gives the excitation profile of the well-known Schottky anomaly, and of course any other two level system in which the excited state is non-degenerate.

If the defect formation is accompanied by a decrease in lattice frequencies in the vicinity of the defect then an entropic component to the excitation must be present [63]. In this case ΔH of defect formation must be replaced by ΔG = ΔH – TΔS, and the defects will be excited more rapidly as temperature increases because of the additional entropy that can then be generated. This is the reason that entropy-rich interstitial defects tend to dominate crystal lattice thermodynamics at temperatures approaching the melting points. They are believed by many to be critically involved in melting [64] (and sub-lattice melting [65]) processes. The entropy associated with the excitation process now contains both the combinatorial term and an additional term due to the new lattice frequencies. Relative to the vibrational entropy of the ground state, there is now an excess entropy that is greater than the combinatorial term which drives the simple two state system to equal state populations as T → ∞. The excess entropy can be evaluated and compared with the pure combinatorial (or configurational) term. This is done in Fig. 7. Up to high levels of excitation (enough to collapse most crystals – however see [65]), Fig. 7 shows, the excess entropy and the configurational entropy remain approximately proportional. This must certainly break down at sufficiently high temperatures because $S_c$ can pass through a maximum when the vibrational excess entropy is large and the high energy state therefore strongly favored by temperature increase.

*(ii) glassformers*. While the above is true of the standard crystal it is to be recognized that the expressions given are identical to those developed in the "bond lattice" or "excitations" model of glass-forming liquids [38b] and its alternative versions [two level models [38a], "quasi-ponctuel defaut" model [38c]. The unapproximated expressions predict sigmoidal excitation profiles [36] that are very similar in form to that found by Sastry et al for the mixed LJ system [37]. Well defined isosbestic points found recently [66] in the vibrational densities of states of a model glassformer (parameterized to resemble o-terphenyl) suggest that excitations models, suitable generalized, may indeed have some validity. Fig. 7 shows that such models have the property that their excess entropies and configurational entropies are proportional, well into the liquid state.

## 6. Other implications of the $S_c/S_{ex} < 1$.

We have above given two examples of liquids in which a large part of the entropy in excess of the crystal ($S_{ex}$) generated above $T_g$ is generated in the vibrational density of states. Although this division of $S_{ex}$ between different sources has been known from glassy state studies [25,26] for more than two decades, it has been either disregarded [7-9, 24, 67, 68] or minimized [69] in discussions of supercooled liquid thermodynamics. The onset of $S_{ex,vib}$ at or near $T_K$ implies a discontinuity in the $S_{vib}$ vs T function (as shown in Fig.1, and Fig 5) which is never represented in standard treatments of supercooled liquid thermodynamics. The vibrational contribution to the liquid properties is almost always based on a smooth continuation into the liquid state of the crystal or glassy vibrational properties [24,67,68]. The revival of interest in this matter is a recent phenomenon [27,36, 70-72], stimulated on one side by the possibility that the source of fragile behavior in liquids might lie in the vibrational density of states [36,70].

It is important to consider its implications a little further in three cases where it will make an important difference to previous conclusions. Firstly, the proposal that the configurational heat capacity is a hyperbolic function of temperature [73,74] is based on the assumption that the vibrational heat capacity can be extrapolated smoothly through $T_g$ with the same slope as for the crystal. If allowance is made for a step change in slope of the vibrational heat capacity at $T_g$, then instead of a $T^{-1}$ dependence, the configurational heat capacity will have a sharper dependence, closer to the $T^{-2}$ dependence required by the Gaussian distribution of configurational microstates ( e.g. the random energy, and random first order, models [39,40]. Support for the random energy model is tantamount to support for a sharp "endpoint" at the Kauzmann temperature [1,3, 39,40] over the more gradual phenomenon implicit in models with binomial densities of states [36,38].

Secondly, the estimates of where the "top of the landscape" (ToL) should lie in temperature, relative to the Kauzmann temperature [69] have been based on assignment of the full change of heat capacity at $T_g$ to the configurational degrees of freedom. Integration of this excess heat capacity, assumed to be hyperbolic in temperature dependence, from $T_K$ to a temperature $T_{ToL}$ at which the entropy $\alpha R$ associated with $\exp(\alpha N)$ basins on the landscape of an N particle system ($\alpha$ close to unity) [3b,75,76] was fully excited, then placed the "ToL temperature" at $1.59T_K$. Subsequent studies of the inherent structure energy profile [37] showed this value to be far too low. A possible underestimate due to some contribution from vibrational excess heat capacity was recognized [69] but the effect was downplayed. If the magnitude of the configurational heat capacity is reduced to 50-80% of the value previously adopted, then the temperature at which the ToL is reached will be greatly increased towards the values indicated by the simulations. A quantitative assessment will be given elsewhere.

Thirdly, in their efforts to assess the size of the cooperatively rearranging group of the Adam-Gibbs theory and its dependence on temperature, Matsuo and co-workers [24] have made a very careful analysis of the components of the glassy state vibrational heat capacity in order to accurately estimate the configurational entropy and its temperature dependence. However these authors have also used a smooth extrapolation through the glass transition rather than allowing for the abrupt change in slope at $T_g$ for the total vibrational entropy that we now see is needed because of the *new* low frequency modes which are generated only above $T_g$ (see figs. 1 and 5). Consequently, as in the second example above, the magnitude and temperature dependence of the configurational heat capacity will have been overestimated, and consequently the size of the cooperative group will need to be re-evaluated.

**7. Conclusions.**

The conclusion of this work is that, for very general reasons associated with the nature of configurational excitation, the total excess entropy of a configurationally excited system, increases in proportion to the configurational component of that entropy, at least till the typical glassformer melting point, 1.5Tg, is reached. This conclusion applies irrespective of whether the system is excited at constant pressure or at constant volume, though in the latter case the proportionality constant may be less than unity. At constant volume the thermodynamic fragility of the liquid is usually less than at constant pressure, due at least in part to the inversion of the vibration frequency effect. For cases where the data are available and appropriately analyzed, the kinetic fragility at constant volume is also smaller than the kinetic fragility at constant pressure. Thus the relation between kinetic fragility and thermodynamic fragility in viscous liquids can be extended at least qualitatively to behavior within a single laboratory system. Abrupt and substantial changes in vibrational heat capacity at $T_g$, which have been overlooked in earlier work, will require revisions in earlier evaluations of the configurational heat capacity and its temperature dependence, as well as of quantities such as the ToL and the CRR that depend on it.

**Acknowledgements.**

This work was supported by the NSF under Solid State Chemistry Grant No. DMR0082535. We thank Srikanth Sastry and Francis Starr for helpful discussions, Wu Xu for assistance with data treatment for Fig. 4(a), and Les Pollard for permission to reproduce Fig. 4(b) from his Ph. D. Thesis (ref. 49(a)).

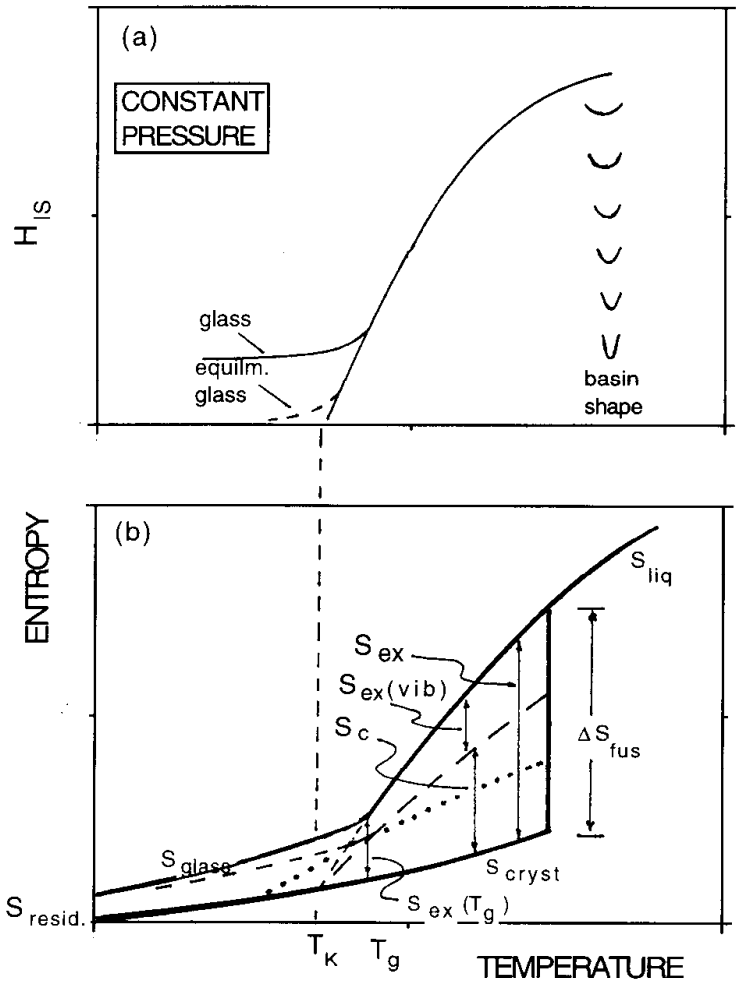

Figure 1. (a) the inherent structure excitation profile dependence on temperature for a typical system.

(b) entropy of a typical crystal, glass and supercooled liquid phases of the same material, showing tendency of all components of the excess entropy of liquid over crystal to vanish at the same temperature, $T_K$. (reproduced from ref. 10 by permission of McMillan Publ.)

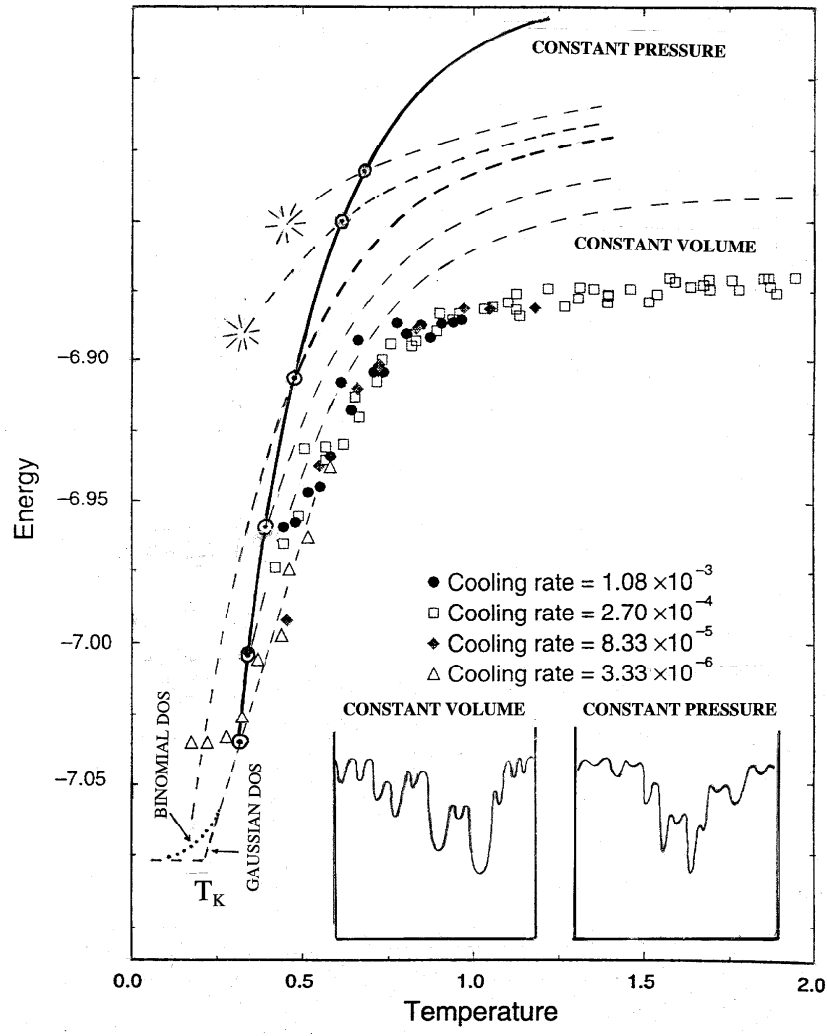

Figure 2. Inherent structure excitation profiles for the mixed LJ system of ref 13, showing the expected behavior of systems of higher fixed volumes, and the steeper profile expected for the system when thermally excited at a constant pressure (the pressure being fixed at the value pertaining at temperature 0.3). The starbursts at the low temperature ends of the upper isochores represent the spinodal limits to supercooling, where the liquid cavitates under extreme negative pressure. The inserts show the features of the energy landscapes that are responsible for these differences. The constant pressure section must be thought of as composed of slices from different constant volume surfaces, the lowest energy sections corresponding to the lowest volumes, such that the pressure of the system remains constant as the temperature is raised and the system point visits successively higher energies. The point of importance is that the basins visited at high temperatures (those near the top of the landscape) are narrower than those at the bottom in the constant volume representation, meaning the vibrational excess entropy decreases with increasing temperature, while in the constant pressure case the opposite is true.

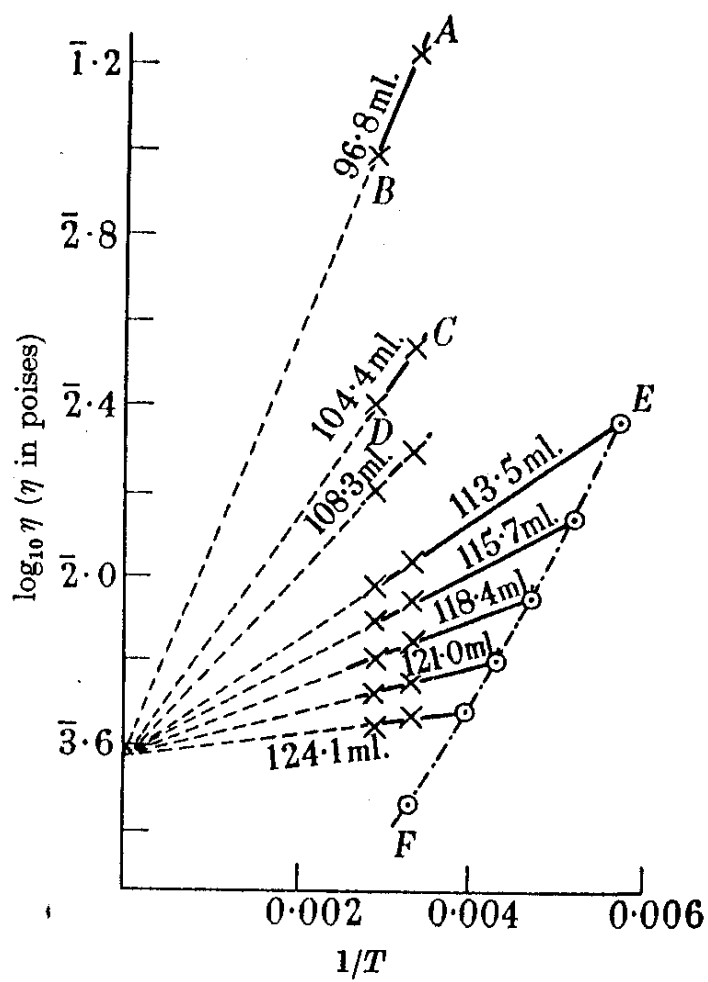

Fig. 3. Viscosity findings of Jobling and Lawrence, ref. 48, for n-hexane studied at constant volumes between 124.1 and 96.8 ml/mol. Note the near vanishing of the activation energy at large constant volumes. The common intersection at $1/T = 0$ for molecular liquids studied at constant pressure ( see e.g. ref. 19) falls between –4 and –5 on this scale. (Reproduced from ref. 48 by permission of the Royal Society of London).

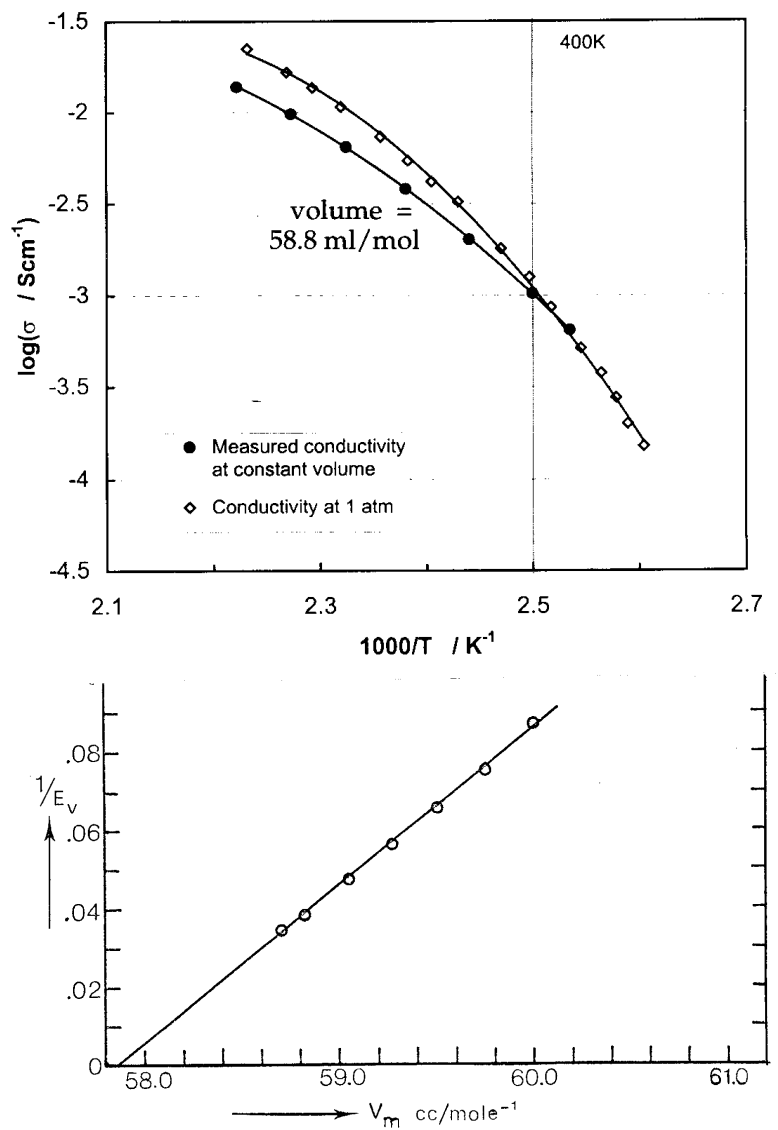

Fig. 4 (a) Comparison of ambient pressure Arrhenius plot for ionic conductivity, with the behavior at a constant volume of 58.8cc/mole. Eq. (1) fit parameters for the two case are given in text and show the constant volume behavior to be 40% less fragile than at constant pressure ($D_{cv}$= 1.4$D_{cp}$). The conductivity in this temperature range is closely coupled to the viscosity.

(b). The volume dependence of the Arrhenius activation energy determined at constant volume using Eq. (8) The figure shows how the constant volume activation energy measured at 1 atm tends to diverge at 57.8 ml/mole. Note that the range of possible volumes left unexplored (by measurements covering only 2.5 orders of magnitude Fig.4 (a)) is small, like the range of energies left unexplored by the simulations of Fig. 2.

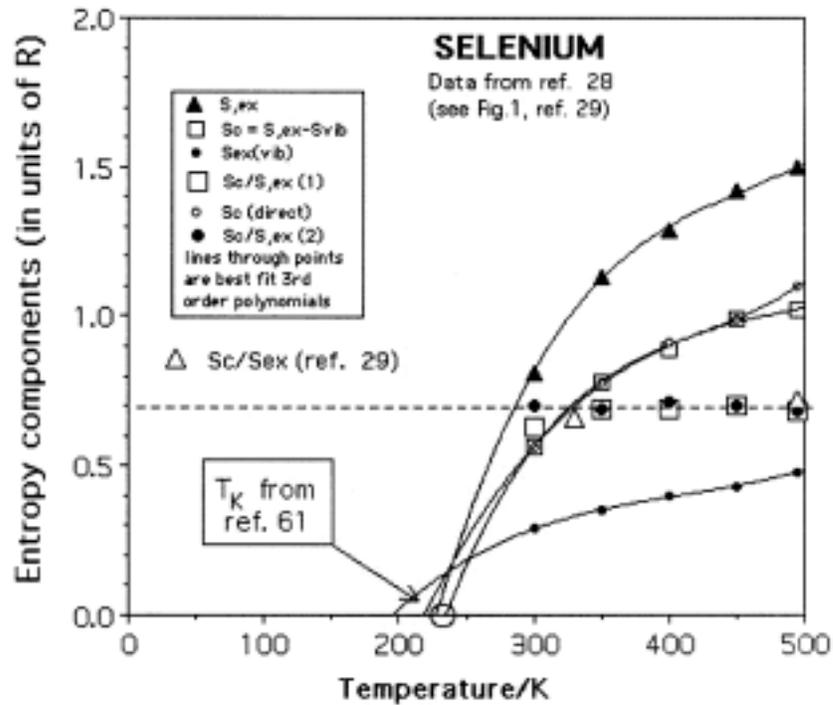

Fig. 5. Dependence on temperature of the excess entropy $S_{ex}$, the configurational entropy $S_c$ and the excess vibrational entropy $S_{ex,vib}$ (over crystal in each case) for the element selenium, showing tendency of each to vanish near the Kauzmann temperature. Large round symbols show the ratio $S_c/S_{ex}$, at each of four temperatures below the melting point. The ratio is seen to maintain an approximately constant value of 0.69. Open triangles are values from ref. 29.

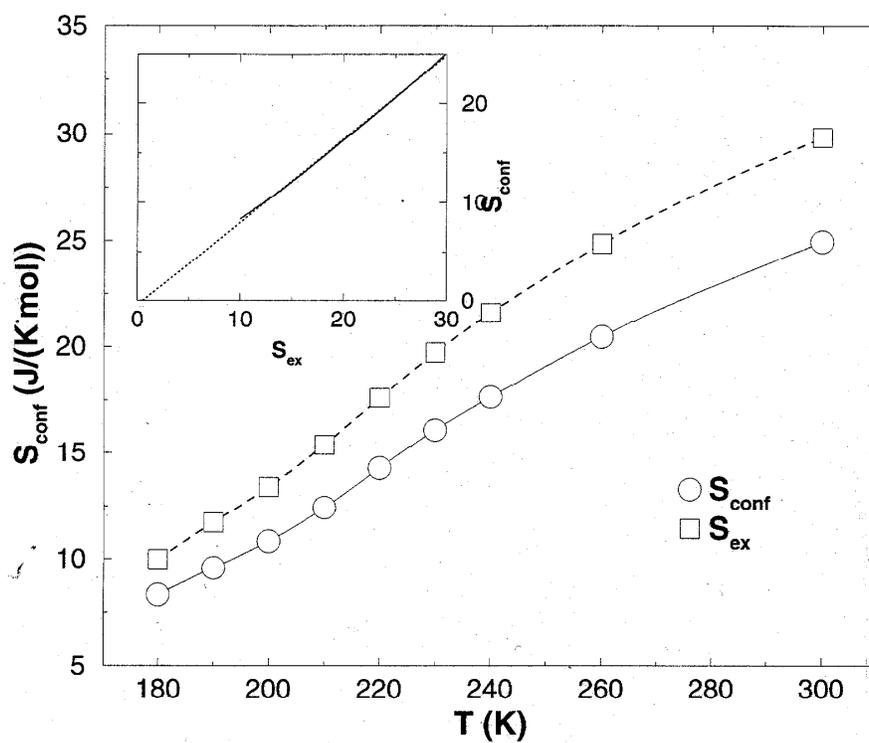

Fig. 6. Variation of $S_c$ and $S_{ex}$ with temperature for the case of SPC-E water, studied by molecular dynamics computer simulation in refs. 2 and X. Insert shows that despite their complex forms, the two quantities remain approximately proportional over the entire range studied. (from ref. 12, by permission of Am. Inst. Phys.).

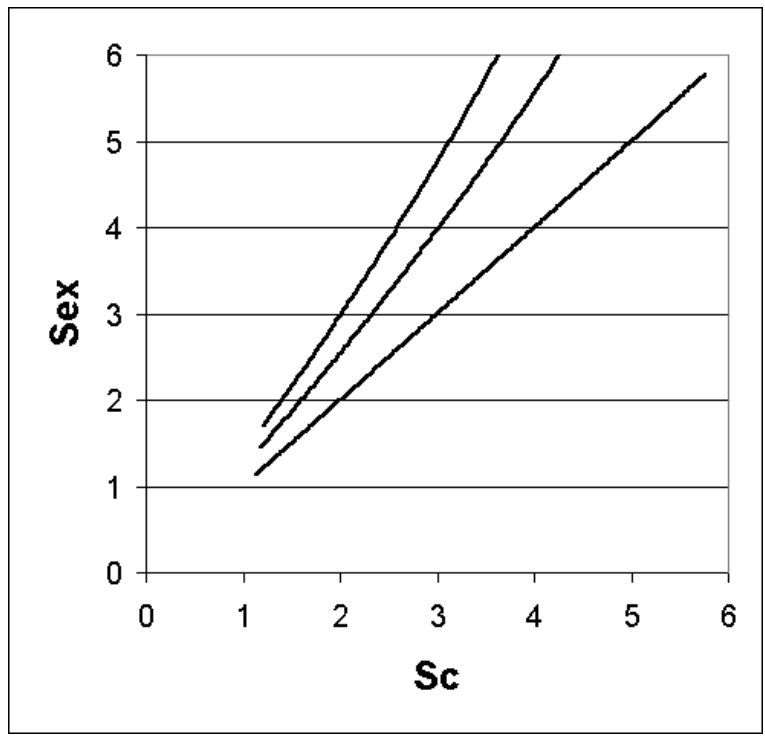

Fig. 7. Total excess entropy, $S_{ex}$, as a function of the configurational component of the excess entropy, $S_c$, for a crystal lattice with defects giving low vibration frequencies near the defects. The model yields a total excess entropy, over the ground state, that is larger than, but approximately proportional to, its configurational (combinatorial) excess entropy for both values of the parameter ($\Delta S^*$) that determines the magnitude of $S_{ex}/S_c$. The lowest line is for $\Delta S^* = 0$.